\documentclass[prd,aps,epsfig,showpacs,twocolumn,floats,superscriptaddress]{revtex4}
\usepackage{mathrsfs}
\usepackage{amssymb,color}
\usepackage{amssymb,graphicx,amsmath}
\begin{document}

\title{A New Integral Equation for the Spheroidal equations in case of m equal 1}
\author{Guihua Tian}
\email{tgh-2000@263.net, tgh20080827@gmail.com}
\author{Shuquan Zhong}
\email{shuqzhong@gmail.com} \affiliation{School of Science, Beijing
University of Posts And Telecommunications, Beijing 100876 China.}

\begin{abstract}
The spheroidal wave functions are investigated  in the case $m=1$.
The integral equation is obtained for them. For the two kinds of
eigenvalues in the differential and corresponding integral
equations, the relation between them are given explicitly. Though
there are already some integral equations for the spheroidal
equations, the relation between their two kinds of eigenvalues is
not known till now. This is the great advantage of our integral
equation, which will provide useful information through the study of
the integral equation. Also an example is given for the special case
, which shows another way to study the eigenvalue problem.

\end{abstract}


\maketitle

\section{Introduction of the spheroidal functions}

The spheroidal wave equations are extension of the ordinary
spherical wave equations. There are many fields where spheroidal
functions play important roles just as the spherical functions do.
So far, in comparison to simpler spherical special functions (the
associated Lengdre's functions ) their properties still are
difficult for study than their
counterpart\cite{flammer}-\cite{li}.

Their differential equations are
\begin{equation} \left[\frac{d}{dx}\left[(1-x^2) \frac{d}{d
x}\right]+E+ \beta^2 x^2
 -\frac{m^{2}}{1-x^2} \right]\Theta=0,\ \label{2}
\end{equation}
where $-1<x<1$.
 This is a kind
of the Sturm-Liouville eigenvalue problem with the natural
conditions that $\Theta $ is finite at the boundaries $x=\pm 1$. The
parameter $E $ can only takes the values $E_0,\ E_1,\dots,
E_n,\dots$, which are called the eigenvalues of the problem, and the
corresponding solutions (the eigenfunctions)
$\Theta_0,\Theta_1,\dots, \Theta_n,\dots$   are called the
spheroidal wave functions \cite{flammer}-\cite{li}.

Under the condition $\beta=0$, they reduce to the Spherical equation
 and  the solutions to the Sturm-Liouville eigenvalue problem are
the associated Legendre-functions $P_n^m(x)$(the spherical
functions) with the eigenvalues $E_n=n(n+1),\ n=m+1,\ m+2,\dots$.
They only have one more term $\beta^2 x^2$ than the spherical ones
(the associated Lendgre's equations). However, the extra term
presents many mathematical difficulties for one to treat the
equations.

Though the spheroidal wave equations are extension of the ordinary
spherical wave functions equations, the difference between this two
kinds of wave funtions are far greater than their
similarity\cite{flammer}.

Usual way to study the spheroidal equations is the perturbation one
resulting in the continued fraction to determine the eigenvalues and
eigenfunctions. Recently, new methods are used to re-investigate the
problems again. The new methods mainly include the perturbation one
in supersymmetry quantum mechanics, which gives rise to many nice
results. Some of the results are the extension of the recurrence
relation of the spherical functions to the spheroidal functions,
which makes one could obtain the excited spheroidal functions from
the ground one. Other results might give new method in their
numerical calculation\cite{tian}. There are also the integral
equations, which provides another way to numerically study the
spheroidal functions. In Ref.\cite{tian1}, the integral equations
are extended to the spin-weighted spheroidal case.

For example, the integral equation for the prolate spheroidal wave
equation is \cite{flammer}-\cite{tian}
\begin{equation}
\Theta(y)=\lambda \int_{-1}^{+1}K(x,y)\Theta(x)dx.\label{8}
\end{equation}
where the kernel $K(x,y)$ is
\begin{equation}
K(x,y)=(1-x^2)^{\frac12m}(1-y^2)^{\frac12m}
\frac{J_{m+\frac12}\left(\bar{\beta}(x-y)\right)}{\left[\bar{\beta}(x-y)\right]^{m+\frac12}}
,\label{27}
\end{equation}
with $\bar{\beta}=i\beta$.  There are two eigenvalues appear in the
differential and the integral equations, that is, the quantities
$E,\ \lambda$ . However, the relation between the eigenvalues $E,\
\lambda$ is unclear \cite{flammer}-\cite{tian}. In this letter, we
will report a new integral equation for the spheroidal equation in
the case of $m=1$.Because the integral equation is derived from the
Green function of the equation,
 the advantage of the new integral equation shows the concise
 relation between the eigenvalues $E,\ \lambda$.

From eq.(\ref{2}) and by the transformation
\begin{equation}
\Theta =\frac{\Psi}{(1-x^2)^{\frac m2}} \label{transform}
\end{equation}
we could obtain the following
\begin{eqnarray}
&&(1-x^2)\frac{d^2\Psi}{dx^2}+2(m-1)x\frac{d\Psi}{dx}\nonumber\\
&&+\left[E-m^2+m+\beta^2- \beta^2(1-x^2)
 \right]\Psi=0.
\end{eqnarray}
The above equation becomes very simple when $m=1$, that
is,\begin{eqnarray}
\frac{d^2\Psi}{dx^2}+\left[\frac{\lambda}{1-x^2}- \beta^2
 \right]\Psi=0, \ -1<x<+1\label{main eq},
\end{eqnarray}
where
\begin{equation}
\lambda=E-m^2+m+\beta^2=E+\beta^2.
\end{equation}
It is easy to find the Green functions for the equations (\ref{main
eq}), that is
\begin{eqnarray}
G(x,\xi)=\frac1{\sinh2\beta}\sinh\beta(1-\xi)\sinh\beta(1+x),\ x<\xi\\
G(x,\xi)= \frac1{\sinh2\beta}\sinh\beta(1-x)\sinh\beta(1+\xi),\
x>\xi
\end{eqnarray}
The Green function $G(x,\xi)$ satisfies  the following
\begin{equation}
\frac{\partial^2 G(x,\xi)}{\partial x^2}- \beta^2G(x,\xi)
 =-\delta(x-\xi)\label{green 0 eq}\end{equation}
 and the boundary conditions
\begin{eqnarray}
G(x,\xi)_{x=-1}=G(x,\xi)_{x=+1}=0
\end{eqnarray}
 Hence the the Sturm-Liouville eigenvalue problem turns into the
 integral equation form:
\begin{eqnarray}
 \Psi(x)&=&\lambda \int_{-1}^{+1}G(x,\xi)\frac{\Psi(\xi)}{1-\xi^2}d\xi\\
 &=&\frac{\lambda}{\sinh2\beta}\bigg[\int_{-1}^{x}\sinh\beta(1-x)\sinh\beta(1+\xi)\Psi(\xi)d\xi\nonumber\\
 &&\ \ \
 +\int_{x}^{1}\sinh\beta(1+x)\sinh\beta(1-\xi)\Psi(\xi)d\xi\bigg]\label{asymmetry integral
 eq.}
\end{eqnarray}
The great advantage lies in that the relation between the integral
eigenvalues $\frac{\lambda}{\sin2\beta}$ and $E$ of the differential
equations for the spheroidal is given explicitly by \begin{eqnarray}
\frac{\lambda}{\sinh2\beta}=\frac{E-m^2+m-\beta^2}{\sinh2\beta}
                                                    \end{eqnarray}
Though the green function $G(x,\xi)$ is symmetry with respect to the
variables $x,\ \xi$, the kernal in Eq.(\ref{asymmetry integral
 eq.}) is not symmetrical at all. Nevertheless, it is easy to make
 the kernal be symmetry. That is, changing $\Psi(x)$ into
 $\hat{\Psi}=\frac{\Psi(x)}{\sqrt{1-x^2}}$, Eq.(\ref{asymmetry integral
 eq.}) becomes
\begin{eqnarray}
\hat{ \Psi}(x)&=&\lambda
\int_{-1}^{+1}\frac{G(x,\xi)}{\sqrt{1-x^2}\sqrt{1-\xi^2}}\hat{
\Psi}(\xi)d\xi,
 \end{eqnarray}
as desired by our requirement. It is well-known that one could
easily to study the integral equations if their kernels are
symmetry.
 Hence, the usual method to solve the integral equations could be
 used to treat the problem here too. We will stop here.

The Green function $G(x,\xi)$ for the spheroidal equations in $m=1$
includes all cases of the parameter $\beta$ as complex number. When
$\beta$ is pure imaginary, the corresponding equation is the prolate
spheroidal equation and the Green function turns out as
\begin{eqnarray}
G(x,\xi)&=& \frac1{\sin2\bar{\beta}}\sin\bar{\beta}(1-\xi)\sin\bar{\beta}(1+x),\ x<\xi\nonumber\\
G(x,\xi)&=&
\frac1{\sin2\bar{\beta}}\sin\bar{\beta}(1-x)\sin\bar{\beta}(1+\xi),\
x>\xi\label{green fun in beta imaginary}
\end{eqnarray}
where $\bar{\beta}=i\beta$ is real. If one supposes the parameter
$\lambda=E+\beta^2=E-\bar{\beta^2}=0$, the parameter $\bar{\beta}$
will stand in the position of the eigenvalues in the
 Sturm-Liouville eigenvalue problem. Of course, the parameter $\beta$ or $\bar{\beta}$
 is no longer a fixed quantity in this case. Notice that the
 case is special because the parameter $\bar{\beta}$ is not a fixed quantity in contrasting with the usual cases.
The original equation becomes \begin{equation}
\left[\frac{d}{dx}\left[(1-x^2) \frac{d}{d
x}\right]+\bar{\beta}^2(1-x^2)
 -\frac{1}{1-x^2} \right]\Theta=0,\ \label{trans eq beta imaginary}
\end{equation}
The Green function in
 Eq.(\ref{green fun in beta imaginary}) will give much information
 about the eigenvalues and eigenfunctions in this special case.
  Now the Green function could be regarded as
 the functions of the parameter $\bar{\beta}$. One could expands
 this Green function in the form \begin{eqnarray}
G(x,\xi)&=&\sum_{n=0}^{\infty}\frac{\Psi_n(x)\Psi_n(\xi)}{\bar{\beta}^2-\bar{\beta}^2_n}
\end{eqnarray}
 The eigenvalues are
 determined by the poles of the Green functions, that is
\begin{eqnarray}
\sin2\bar{\beta}=0.
\end{eqnarray}
Hence, $\bar{\beta}^2=\frac{n^2\pi^2}4,\ n=1,2,\cdots,$  are the
eigenvalues, and the residues of the corresponding pole are
\begin{eqnarray}
\frac{\Psi_n(x)\Psi_n(\xi)}{\bar{\beta}_n^2}&=&\bigg[
\frac{G(x,\xi)}{\bar{\beta}^2-\bar{\beta}^2_n}\bigg]_{\bar{\beta}=\bar{\beta}_n}\nonumber\\
&=& \frac1{2\bar{\beta}_n}\sin\frac{n\pi}{2}\xi\sin\frac{n\pi}{2}x,\
n=2,4,6\cdots,\label{green fun expansion n is even}
\end{eqnarray}
and
\begin{eqnarray}
\frac{\Psi_n(x)\Psi_n(\xi)}{\bar{\beta}_n^2}&=&\bigg[
\frac{G(x,\xi)}{\bar{\beta}^2-\bar{\beta}^2_n}\bigg]_{\bar{\beta}=\bar{\beta}_n}\nonumber\\
&=& \frac1{2\bar{\beta}_n}\cos\frac{n\pi}{2}\xi\cos\frac{n\pi}{2}x,\
n=1,3,5\cdots.\label{green fun expansion n is odd}
\end{eqnarray}
the nth eigenfunction is\begin{eqnarray}
\Psi_n(x)&=&\sqrt{\frac{\bar{\beta}_n}{2}}\sin\frac{n\pi}{2}x,\
n=2,4,6,\cdots,\\
\Psi_n(x)&=&\sqrt{\frac{\bar{\beta}_n}{2}}\cos\frac{n\pi}{2}x,\
n=1,3,5,\cdots
\end{eqnarray}
Except for the normalization constants, these results are the same
as those in Ref.\cite{flammer}, though they are derived from the
different way. As stated in Ref.\cite{flammer}, the function
\begin{eqnarray}\Theta_n=\frac{\Psi_n(x)}{(1-x^2)^{\frac
12}}=\sqrt{\frac{\bar{\beta}_n}{2}}
\frac{\sin\frac{n\pi}{2}x}{(1-x^2)^{\frac 12}}\end{eqnarray} is one
of the eigenfunctions for the fixed parameter
$\bar{\beta}=\frac{n\pi}2,\ n=2,4,6,\cdots $ of the original
equation (\ref{2}) in case $m=1$, so does
\begin{eqnarray}\Theta_n=\frac{\Psi_n(x)}{(1-x^2)^{\frac
12}}=\sqrt{\frac{\bar{\beta}_n}{2}}
\frac{\cos\frac{n\pi}{2}x}{(1-x^2)^{\frac 12}}\end{eqnarray} for the
fixed parameter $\bar{\beta}=\frac{n\pi}2,\ n=1,3,5,\cdots $.

The above example just provides some clues on the connection between
the Green function and the solutions to the corresponding the
 Sturm-Liouville eigenvalue problem. If the Green function is the
 one corresponding with the parameter $\lambda \ne 0$, they will
 more useful than just giving the integral equation. However, one could not obtain
 directly the information on the eigenvalues and eigenfunctions from the the Green function
  corresponding with the parameter $\lambda = 0$ . In this
  situation, the useful information could be obtained through the
  study on the integral equation. Here the Green function $G(x,\xi)$
 satisfies Eq.(\ref{green 0 eq}), rather than the following
\begin{equation}
\frac{\partial^2 \bar{G}(x,\xi)}{\partial x^2}+
\left[\frac{\lambda}{1-x^2}+ \beta^2
 \right]\bar{G}(x,\xi)
 =-\delta(x-\xi)\label{green   eq}\end{equation}
This Green function $\bar{G}(x,\xi)$ is connected with the
eigenfunctions $\Psi_n(x)$ by
\begin{eqnarray}
\bar{G}(x,\xi)=-\sum_{n=0}^{\infty}\frac{\Psi(x)\Psi(\xi)}{\lambda-\lambda_n}
=-\sum_{n=0}^{\infty}\frac{\Psi_n(x)\Psi_n(\xi)}{E-E_n}.
\end{eqnarray}
Our Green function $G(x,\xi)$ related to $\bar{G}(x,\xi)$ by
\begin{eqnarray}
G(x,\xi)&=&\bar{G}(x,\xi)_{\lambda=0}=\sum_{n=0}^{\infty}\frac{\Psi_n(x)\Psi_n(\xi)}{\lambda_n}\\
&=&\sum_{n=0}^{\infty}\frac{\Psi_n(x)\Psi_n(\xi)}{E_n-\beta^2}
\end{eqnarray}
Of course, $\bar{G}(x,\xi)$ contains much more useful information on
the eigenvalues and eigenfunctions than that of $G(x,\xi)$, but it
is much harder to obtain. Even it is inferior to $\bar{G}(x,\xi)$,
$G(x,\xi)$ still could provide useful information through the
integral equation, which will be our further study.
\section*{Acknowledgements}
This work was supported in part by the National Science Foundation
of China  under grant No.10875018.

\end{document}